\documentclass[twocolumn,preprintnumbers,amsmath,amssymb]{revtex4}  

\usepackage{graphicx}
\usepackage{dcolumn}
\usepackage{bm}
\usepackage{multirow}
\usepackage{color}
\usepackage{hyperref}

\begin{document}

\title{Nuclear magnetization in gallium arsenide quantum dots at zero magnetic field}

\author{G.~Sallen$^1$}
\author{S.~Kunz$^1$}
\author{T.~Amand$^1$}
\author{L.~Bouet$^1$}
\author{T.~Kuroda$^2$}
\author{T.~Mano$^2$}
\author{D. Paget$^3$}
\author{O.~Krebs$^4$}
\author{X.~Marie$^1$}
\author{K.~Sakoda$^3$}
\author{B.~Urbaszek$^1$}

\affiliation{%
$^1$Universit\'e de Toulouse, INSA-CNRS-UPS, LPCNO, 135 Av. Rangueil, 31077 Toulouse, France}

\affiliation{%
$^2$National Institute for Material Science, Namiki 1-1, Tsukuba 305-0044, Japan}

\affiliation{%
$^3$LPMC, Ecole Polytechnique, CNRS, 91128 Palaiseau, France}

\affiliation{%
$^4$CNRS Laboratoire de Photonique et de Nanostructures, Route de Nozay, 91460 Marcoussis, France}


\begin{abstract}
Optical and electrical control of the nuclear spin system allows enhancing the sensitivity of NMR applications and spin-based information storage and processing. Dynamic nuclear polarization in semiconductors is commonly achieved in the presence of a stabilizing external magnetic field. Here we report efficient optical pumping of nuclear spins at zero magnetic field in strain free GaAs quantum dots. The strong interaction of a single, optically injected electron spin with the nuclear spins acts as a stabilizing, effective magnetic field (Knight field) on the nuclei. We optically tune the Knight field amplitude and direction. In combination with a small transverse magnetic field, we are able to control the longitudinal and transverse component of the nuclear spin polarization in the absence of lattice strain i.e. nuclear quadrupole effects, as reproduced by our model calculations.

\end{abstract}

\maketitle
\begin{figure}
\includegraphics[width=0.48\textwidth]{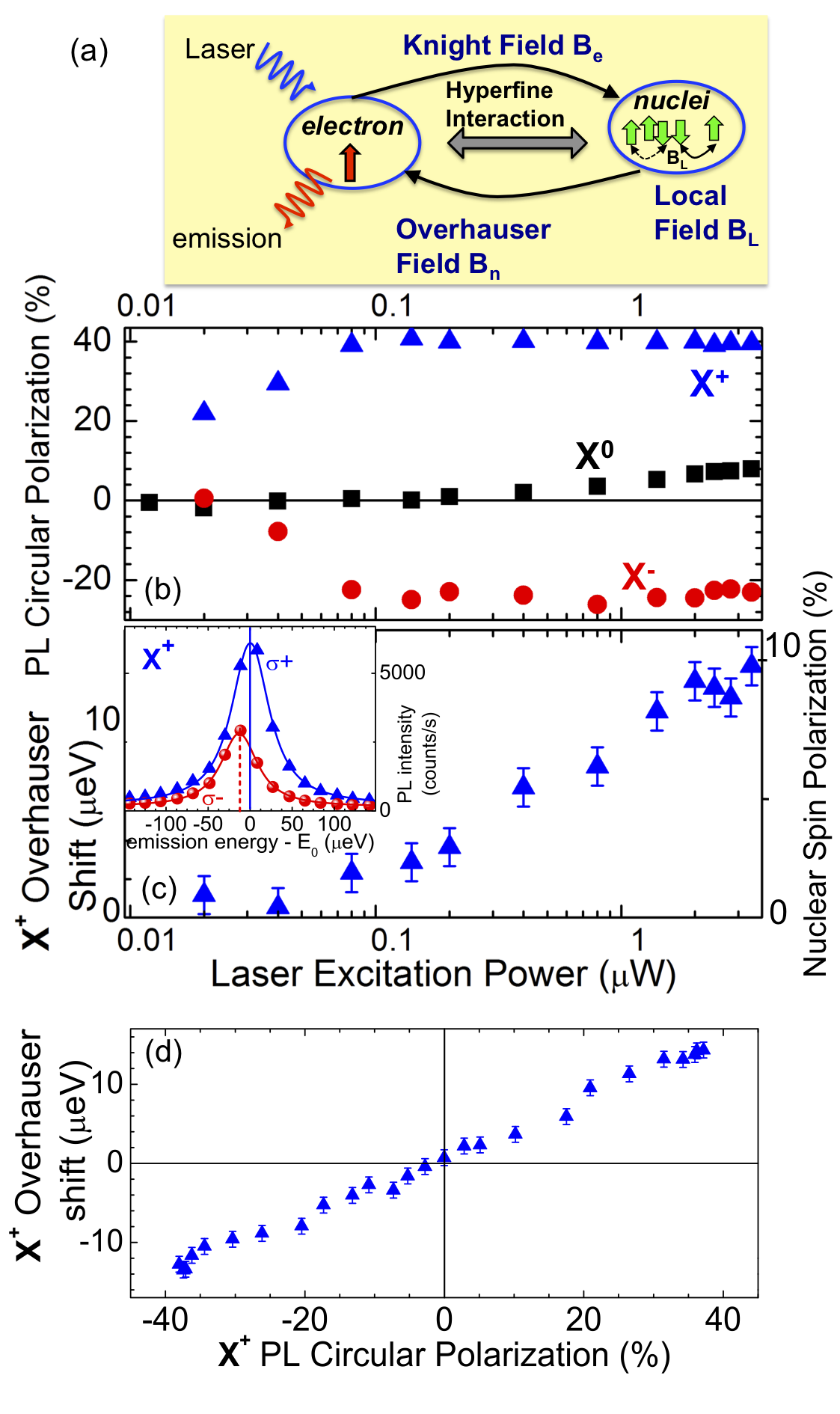}
\caption{\label{fig:fig1} \textbf{Dynamic nuclear polarization at zero applied field} (a) Hyperfine interaction between a single electron and the nuclear spins of the quantum dot lattice.(b) Circular polarization degree of emission (error $\pm$ 1\% included in symbol height) of neutral exciton X$^0$ (squares) and charged excitons X$^+$ (triangles) and X$^-$ (circles) as a function of excitation laser power using $\sigma^+$ polarization (c) Energy difference between $\sigma^+$ and $\sigma^-$ polarized emission for X$^+$: Overhauser shift $\propto$ average nuclear spin $\langle I_z \rangle$ Inset: X$^+$PL spectra $\sigma^+$ (triangles) and $\sigma^-$(circles) polarized at max. laser power. (d) Overhauser shift as a function of the measured X$^+$ PL polarization with varying laser excitation polarization at constant power of 2~$\mu$W.
}
\end{figure}

Investigation of dynamic nuclear polarization (DNP) effects in bulk semiconductors \cite{Meier:1984a,King:2012a}, in quantum wells \cite{Kalevich:1990a,Salis:2001a}, and in quantum dots \cite{Gammon:1997a,Urbaszek:2013a} under circularly-polarized light excitation has been a very active field of research for more than 30 years, both from a fundamental point of view and for potential applications in quantum computing and NMR imaging\cite{Reimer:2010a,Kaur:2010a}. These effects, predicted originally for metals \cite{Overhauser:1953a} manifest themselves by giant hyperfine fields of nuclear origin experienced by localized electrons \cite{Paget:1977a} and can influence their spin polarization in the same way as an external magnetic field.   
The concept of a temperature among a nuclear spin system, although originally developed for insulators \cite{Abragam:1958a,Abragam:1961a}, has been extremely fruitful for explaining most of these results. It opens the possibility to manipulate both the magnitude \cite{Berkovits:1978a,Dyakonov:1975a} and the orientation \cite{Giri:2012a} of the nuclear magnetization, and therefore of the large hyperfine nuclear field $B_n$, in a small external magnetic field, of the order of the local field $B_L$ which characterizes nuclear spin-spin interactions, see Fig.\ref{fig:fig1}a. The concept of nuclear spin temperature implies that no nuclear field can build up in a magnetic  field smaller than $B_L$, where $B_L\approx 0.15$~mT in bulk GaAs \cite{Paget:1977a}.\\
\indent Recently, the possibility to measure  the nuclear spin polarization degree in a single dot with an extremely high precision \cite{Lai:2006a} has led to the observation of DNP in highly strained InGaAs and InP semiconductor quantum dots at zero magnetic field \cite{Lai:2006a,Moskalenko:2009a,Dzhioev:2007a,Belhadj:2009a,Oulton:2007a} and the nuclear spin polarization remained stable in the absence of any electron spin, after removal of the light excitation \cite{Maletinsky:2009a} . All these measurements were performed in quantum dots formed in the Stranski-Krastanov (SK) growth mode and the inevitable strain in these structures creates nuclear quadrupole splittings corresponding to effective internal magnetic fields of tens to hundreds of mT \cite{Chekhovich:2012a,Bulutay:2012a}. These quadrupolar effects (and not as initially thought the Knight field \cite{Lai:2006a,Urbaszek:2013a}) have been put forward as being responsible for the stabilization of nuclear magnetization at zero field.  \cite{Dyakonov:2008a,Dzhioev:2007a,Krebs:2010a,Maletinsky:2009a}. \\ 
\indent Here we report efficient optical initialization of a nuclear spin polarization of $\sim$10 \% in strain free, isolated GaAs quantum dots \cite{Mano:2010a} at zero applied magnetic field. We show that even in the absence of static nuclear quadrupole effects the concept of nuclear spin temperature is not applicable. Thus zero field DNP is possible with a mean electron spin polarization as low as 10\%. This strongly anomalous behavior originates from the very large Knight field $B_e$, the effective field experienced by the nuclei in the presence of a spin polarized electron \cite{Knight:1949a}. 
The strong impact of the large Knight field on the nuclear magnetization amplitude and direction is demonstrated. We show that strong Knight field inhomogeneities between one nucleus and its nearest neighbor prevent both the establishment of a temperature and of spin diffusion among the nuclear spins throughout the dot. \\
\indent Our findings suggest that zero field DNP can occur in other spintronic devices with highly polarized electrons such as gate defined quantum dots \cite{Kobayashi:2011a,Petersen:2013a,Hanson:2007a,Bluhm:2010b,Hermelin:2011a} and ferromagnet/semiconductor hybrid structures \cite{Shiogai:2012a,Salis:2009a}. Controlling nuclear spin polarization is also important for semiconductor based quantum emitters, usually operating at zero applied magnetic fields, as the Overhauser field and its fluctuations will alter the polarization basis of the emitted photons in entanglement experiments  \cite{Kuroda:2013a,Stevenson:2011a}.\\

\textbf{RESULTS. Dynamic Nuclear Polarization at zero applied magnetic field}   

One of the outstanding features of quantum dots is the possibility to achieve DNP in zero external magnetic fields \cite{Lai:2006a,Moskalenko:2009a,Dzhioev:2007a,Belhadj:2009a,Oulton:2007a}. In reference \cite{Maletinsky:2009a} DNP was first constructed via optical pumping in the presence of a well defined electron spin in a charge tunable InGaAs quantum dot. Subsequently the electron was removed from the dot and the nuclear spin polarization remained stable in the absence of any electron spin. In this case the Knight field is also absent and strong nuclear quadrupole effects in strained dots with alloy disorder  have been put forward as the main source for stabilising the nuclear spin system \cite{Dzhioev:2007a}. \\
\indent Here, as shown in Fig.\ref{fig:fig1}c and d, we provide first experimental evidence for substantial DNP in zero external field, in the absence of nuclear quadrupole effects. All the results shown in the present work were taken  in \textit{unstrained} GaAs droplet dots with negligible alloy intermixing \cite{Keizer:2010a}. Our sample is subject to charge fluctuations \cite{Mano:2010a,Sallen:2011a}. During the 1s photoluminescence (PL) integration time we observe charged (X$^+$ and X$^-$) and neutral (X$^0$) emission of a single quantum dot. The  X$^+$ (2 holes in a spin singlet, 1 electron) with a well orientated electron spin can polarize the nuclei via the Fermi-contact hyperfine interaction. Note that the PL polarization of the X$^-$ (2 electrons in a spin singlet, 1 hole) is negative in Fig.\ref{fig:fig1}b , i.e. cross-polarized with respect to the excitation laser \cite{Bracker:2005a,Laurent:2006a,Cortez:2002a,Verbin:2012a}. The resident conduction electron left behind after X$^-$ recombination has therefore the same orientation as the conduction electron which exists during the radiative lifetime of the X$^+$. The DNP resulting from this single electron (that will eventually tunnel out of the dot) and the DNP built up during the X$^+$ radiative lifetime have the same direction. The contribution of the X$^+$ and X$^-$ exciton to the nuclear polarization are thus additive. We can neglect the small contribution of the hole spin hyperfine interaction in a first approximation \cite{Chekhovich:2013a,Fischer:2008a,eble09}. The neutral X$^0$ will feel the nuclear field created by the charged excitons \cite{Belhadj:2009a}. In the remainder of the paper we concentrate on the X$^+$ exciton and hence focus on the interaction of the unpaired electron with the nuclei during the radiative lifetime, the mean electron spin is related to the X$^+$ PL circular polarization degree $\rho$ simply as $\langle S_z \rangle = -\rho/2$ (see Methods).\\
\indent In the absence of any internal or applied magnetic field, the X$^+$ transition shows a single line. As we increase the laser excitation power, a clear splitting between the circularly $\sigma^+$ and $\sigma^-$ polarized components emerges (Overhauser shift), as seen in Fig.\ref{fig:fig1}c and inset. As the laser power is increased, the electron spin is transferred more and more efficiently to the nuclear spins via the Overhauser effect \cite{Overhauser:1953a}. We measure an Overhauser shift of up to 16~$\mu$eV (see Fig.\ref{fig:fig1}c and d), which corresponds to a nuclear polarization of 12\%, taking into account that the Overhauser shift for 100\% nuclear polarization in GaAs is 135~$\mu$eV \cite{Gammon:1997a}. In Fig.\ref{fig:fig1}d we effectively tune the injected electron spin polarization via the excitation laser polarization $P^L_c$ as $\langle S_z\rangle^0 \propto -\frac{P^L_c}{2}$ and we observe zero field DNP for an electron spin polarization as low as 10\%. This average spin is low compared to the electron spin injected electrically in hybrid ferromagnet/GaAs devices \cite{Shiogai:2012a,Salis:2009a}. In these systems zero field DNP has not been clearly identified, in part due to permanent stray magnetic fields in the device \cite{Vandorpe:2005a}.

\begin{figure}
\includegraphics[width=0.5\textwidth]{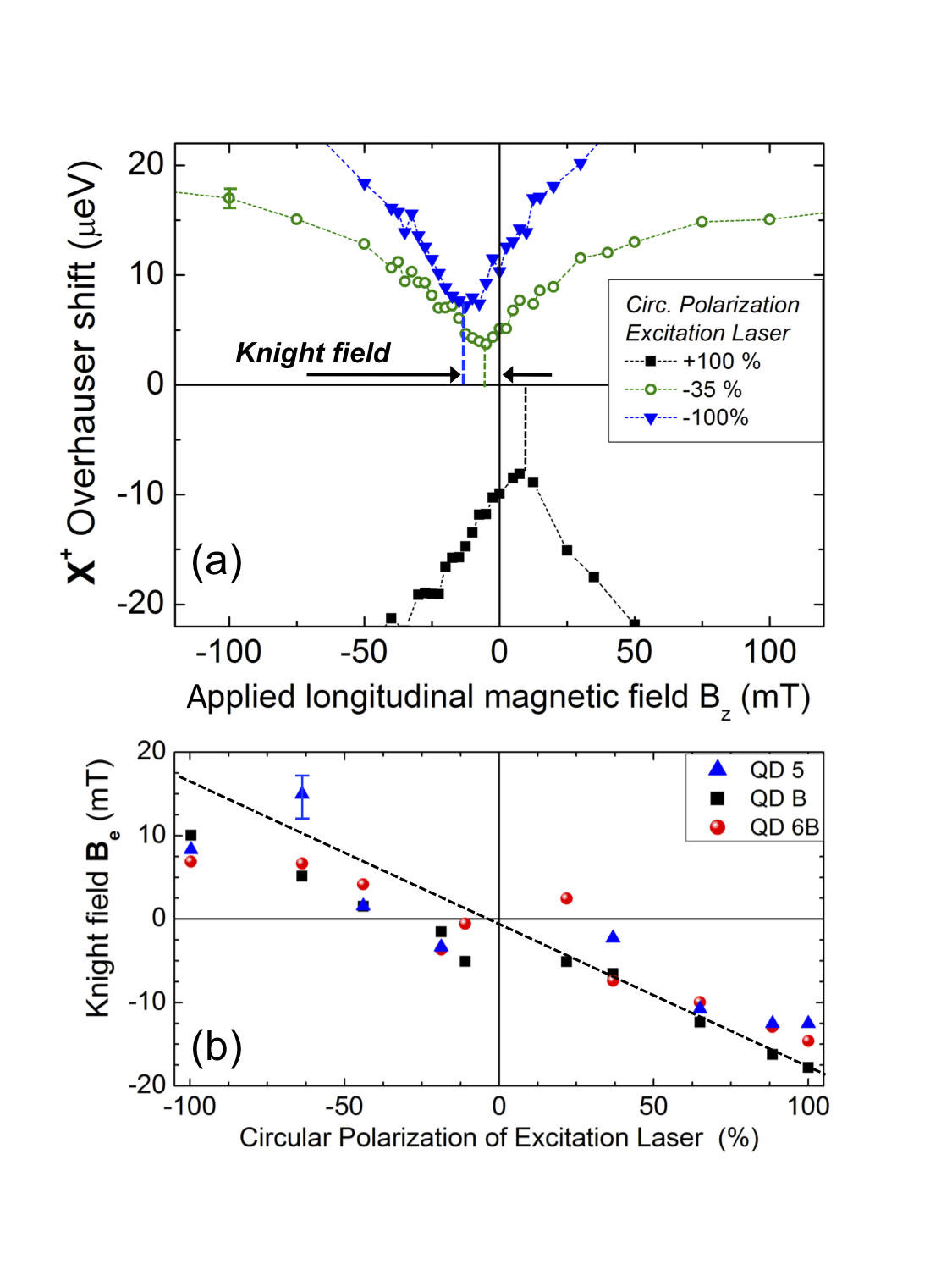}
\caption{\label{fig:fig2} \textbf{Knight field determination and tuning}  (a) \textbf{X$^+$} Overhauser shift as a function of the applied field $B_z$. circles (squares) for $\sigma^-$($\sigma^+$) laser polarization, triangles for elliptically polarized laser excitation. The magnetic field $B_z$ for which this dip occurs, is a measure of the Knight field amplitude. (d) Knight field amplitudes extracted by fitting Gaussians to data in (a) for 3 typical dots, demonstrating Knight field tuning. The data shows a roughly linear dependence as $B_e\propto \langle S\rangle \propto (-P^{L}_c/2)$, typical error bars are indicated.
}
\end{figure}

\begin{figure*}[!]
\includegraphics[width=0.95\textwidth]{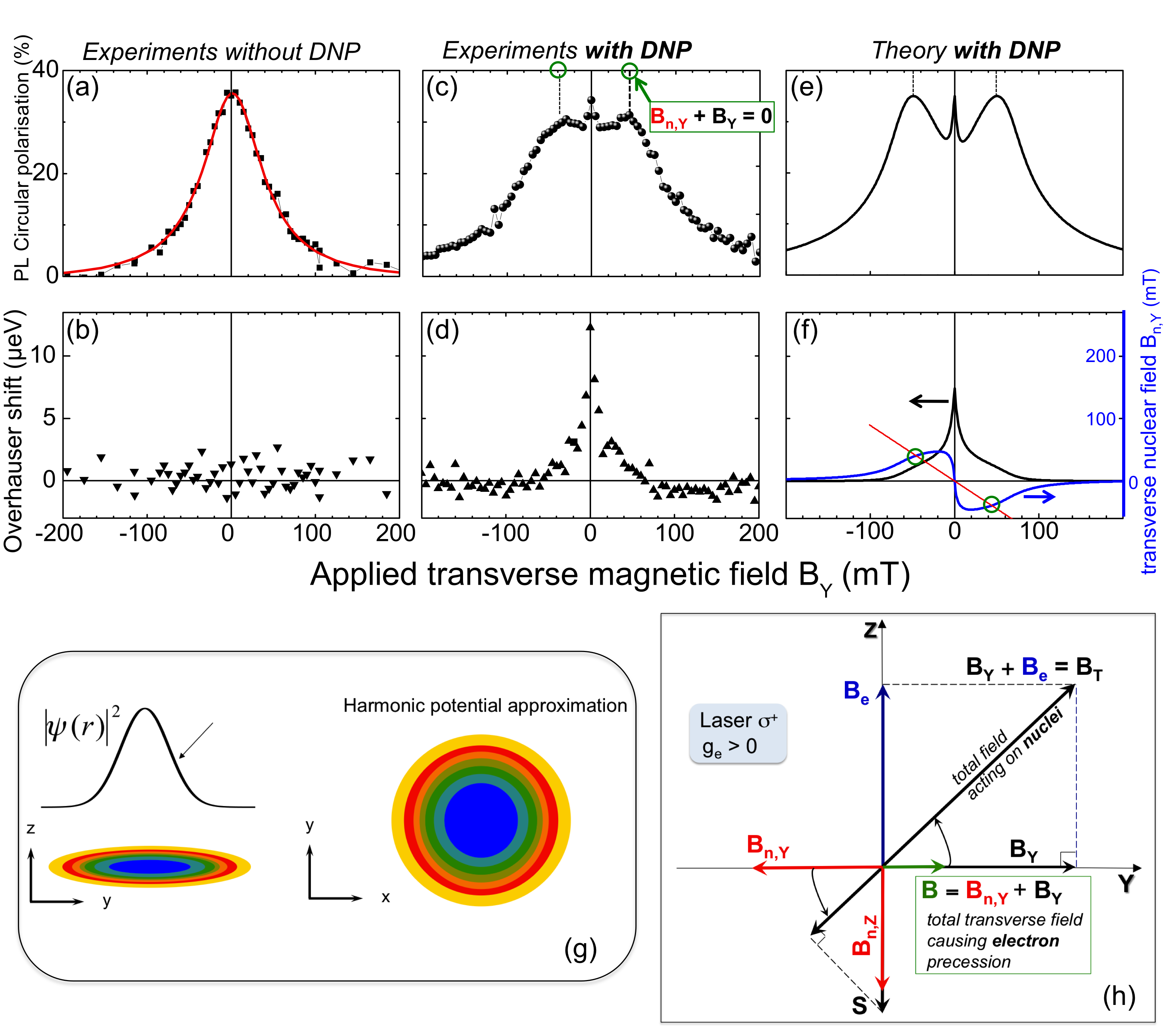}
\caption{\label{fig:fig3} \textbf{Hanle measurements under optical pumping conditions: tilting the nuclear magnetization} (a) Circular polarisation degree of the PL emission as a function of the applied transverse magnetic field. Measurement (black circles) at low laser power (13~nW). Red line: Lorentzian fit. (b)  Overhauser shift $\simeq0$ i.e. there is no measurable DNP.  (c) Same measurement as (a) but using higher laser excitation power (1.2 $\mu$W). (d) strong Overhauser shift under high excitation power. (e) Calculated Hanle depolarization curve using Eq.\ref{eq:Ptheo} - see Methods (f) calculated longitudinal Overhauser field (black line) using Eq. \ref{eq:Bnz}. and transverse nuclear field (blue line). The compensation points where $B_{n,Y}+B_Y=0$ are marked by green circles. Red line: $g_{e,\perp} \mu_B B_Y$ (g) Spin temperature within ellipsoids of equal electron probability density in the dot (h) Effective and applied magnetic fields acting on electron and nuclear spin system - see Methods.
}
\end{figure*}

\textbf{Knight field tuning.}  
In the absence of strain (i.e. no nuclear quadrupole effects), DNP at zero magnetic field could be enabled by a strong Knight field. The time averaged Knight field of an electron in a state $\psi(\vec{r}_j)$ acting
on one specific nucleus $j$ given by \cite{Dyakonov:2008a}:
\begin{equation}
\label{eq:Ber}
\bm{B}_{e,j}\equiv \Gamma_t \frac{\nu_0 A^j}{g_N \mu_N} \vert\psi(\vec{r}_j)\vert^2 \langle \hat{\bm{S}} \rangle \equiv \Gamma_t b_e^j(\vec{r}_j)\langle \hat{\bm{S}} \rangle
\end{equation}
where $\nu_0$ is the two atom unit cell volume, $\vec{r}_j$ is the
position of nucleus $j$, $A^j$ is the hyperfine interaction constant $\simeq 50 \mu$eV on average for Ga and As. $g_N$ and $\mu_N$ are the nuclear g-factor and magneton, respectively. The mean electron spin $\langle \hat{\bm{S}} \rangle$ averaged over the X$^+$ lifetime is expressed in units of $\hbar$. $0\leq \Gamma_t \leq 1$ represents the fraction of time that the dot is occupied by an electron, so $\Gamma_t$ increases with laser power until saturation underlining that the Knight field is zero in the absence of electrons. The time and spatially averaged Knight field determined in single dot spectroscopy can be written as:
\begin{equation}
\langle \bm{B}_e(\bm{r})\rangle _{QD}=\Gamma_f \Gamma_t b^j_e(0)\langle \hat{\bm{S}}\rangle
\label{eq:Berav}
\end{equation}
where $b^j_e(0)\langle S \rangle$ is the Knight field in the centre of the dot where the electron probability density is at a maximum. The form factor for a dot with harmonic confinement is $\Gamma_f=\frac{1}{2^{3/2}}\simeq 0.35$. For the case of electrons localised near donors i.e. a Coulomb potential, the form factor is about 3 times smaller as $\Gamma_f \simeq \frac{1}{e^{2}}\simeq 0.135$. This is one of the reasons why the Knight fields reported here in dots are stronger than those reported in bulk GaAs for electrons localized on donors \cite{Paget:1977a} and are likely to be larger than the local field $B_L$, thus stabilizing the nuclear spins, see Supplementary Information. This would imply that it is possible to compensate the Knight field by a longitudinal magnetic field $B_Z$ in the mT range  \cite{Lai:2006a,Paget:1977a}, such that when $\langle B_e\rangle_{QD}=-B_Z$ the nuclei experience a zero total field and that nuclear depolarization due to the local field $B_L$ becomes possible. One expects the global minimum of the measured DNP as a function of $B_Z$ to appear at this point and not at zero field. \\
\indent This is indeed what we find in the experiments, see Fig.\ref{fig:fig2}a.  When exciting the dot with a $\sigma^-$ polarized laser the Overhauser shift passes through a minimum at $B_{Z}\approx -10$~mT. When changing the helicity of the light polarization, the minimum in absolute value is observed at $B_{Z}\approx +10$~mT, with intermediate values for elliptical polarization (circles in Fig.\ref{fig:fig2}a). Electron spin relaxation in these small longitudinal applied fields is negligible. The Knight field depends on the average electron spin $\langle \hat{\bm{S}} \rangle$, see Eq.\ref{eq:Ber} and \ref{eq:Berav}. The amplitude and direction of the mean electron spin in the X$^+$ are controlled by the polarization of the excitation laser.  This allows for Knight field tuning \cite{Makhonin:2010a}. The dependence of the applied field $B_Z$ at which the minimum Overhauser shift is observed as a function of injected mean electron spin (i.e. laser polarization) is roughly  linear in Fig. \ref{fig:fig2}b. This is a strong indication that the Knight field compensation indeed determines the global minimum of the DNP as a function of $B_Z$. With respect to the work on zero field DNP in strained InGaAs dots by \textcite{Lai:2006a}, the Knight field amplitudes reported here of up to $\langle B_e \rangle_{QD}^{max}=18$~mT are one order of magnitude higher. The initial interpretation of \textcite{Lai:2006a} was based on the Knight field stabilization of zero field DNP, this interpretation has subsequently evolved as nuclear quadrupole effects seem to be dominant in strained dots \cite{Maletinsky:2009a,Urbaszek:2013a}. \\
\textbf{Knight field inhomogeneity.} The measured nuclear spin polarization i.e. the Overhauser shift does not drop abruptly to zero in Fig.\ref{fig:fig2}a. The Knight field experienced by each individual nucleus will depend on its position in the quantum dot as the electron wavefunction density decreases away from the dot centre. Due to this spatial inhomogeneity of the Knight field the exact compensation by the external field occurs only for fraction of the total number of nuclei in the dot. The other nuclei can still be dynamically polarized and therefore contribute to the remaining Overhauser shift.\\ 
\indent An immediate consequence of the large spatial inhomogeneities of the Knight field is that, contrary to the case of electrons localised on donors in bulk GaAs \cite{Paget:1977a}, the concept  a nuclear spin temperature \cite{Abragam:1961a} among the whole nuclear spin system is not applicable, as detailed in the Supplementary Information. Based on our measurements (see Fig.\ref{fig:fig2}), we estimate that the Knight field gradient between two lattice sites of the same nuclear species in a typical dot in our sample is about one order of magnitude larger than $B_L$ (except close to the dot centre and the edges where $\overrightarrow{\nabla}_r|\psi(r)|^2$ vanishes). As a result the nuclear spin diffusion from the dot centre to the outer layers is expected to be suppressed and no global spin temperature can be established in the dot. We therefore assume that only nuclei located on ellipsoids with equal electron wavefunction density can thermalise among themselves, as indicated in Fig.\ref{fig:fig3}g, so that no nuclear spin coherence appears (in the cw regime). As a consequence, the average nuclear spin polarization in a given shell will be collinear with the total field experienced by the nuclei, see Fig.\ref{fig:fig3}h. It is important to underline that despite the apparent Knight field inhomogeneity, the Knight field $B_e$ stabilization of the nuclear spin system at zero field is very efficient and nuclear spin polarization $>10\%$ can be achieved, equivalent to Overhauser fields $B_n$ of several hundred mT.

\textbf{Electron depolarization in applied transverse magnetic fields: Hanle experiments.}  In strained InGaAs dots in the presence of DNP, measurements of the electron depolarization in a transverse magnetic field (Voigt geometry) have led to the discovery of the anomalous Hanle effect \cite{Krebs:2010a}. In these experiments a stable electron spin projection onto the z-axis was observed despite the application of strong transverse fields $B_Y\simeq1$~T \cite{Nilsson:2013a}. Nuclear quadrupole effects are thought to be at the origin of this stabilization as a transverse Overhauser field is constructed. This results in a total magnetic field seen by the electron that is close to zero and no spin precession occurs. Our aim is to verify if the anomalous Hanle effect is also observed in strain free dots or if we can simply tilt the substantial Overhauser field generated at zero applied field through the subsequent application of a transverse field in the mT range \cite{Paget:1977a}.\\
\indent For Hanle measurements, two situations have to be distinguished: with and without optical carrier spin pumping. In the absence of strong carrier spin pumping, and hence nuclear spin pumping, the only field acting on the spin is the applied magnetic field $B_Y$ and a Lorentzian shaped electron depolarization curve is observed \cite{Meier:1984a,Bracker:2005a} as in Fig.\ref{fig:fig3}a in the \textit{absence} of DNP. This is achieved by lowering the laser power to such an extent, that the Overhauser shift completely disappears (Fig.\ref{fig:fig3}b). In Fig.\ref{fig:fig3}a we plot the circular  polarization degree of the X$^+$ PL as a function of the applied transverse magnetic field $B_{Y}$. Having determined the $g_{e,\bot}$ factor beforehand \cite{Durnev:2013a}, fitting the Hanle curve with a Lorentzian allows to extract the electron spin lifetime $\tau_s^*$ as $\Delta B=\hbar/(g_{e,\bot}|\mu_B|\tau_s^*)=43$~mT. The obtained value of $\tau_s^*=350$~ps corresponds roughly to the radiative lifetime of the X$^+$ in these structures \cite{Kuroda:2013a}, which means that the electron spin is stable during the radiative lifetime.\\
\indent In Fig.\ref{fig:fig3}c we repeated the same measurements but with a higher laser excitation power i.e. more efficient carrier and nuclear spin pumping.
This results in a strong Knight field (as we have increased the filling factor $\Gamma_t$ in Eq.\ref{eq:Ber}) and strong DNP as confirmed by a substantial Overhauser shift (Fig.\ref{fig:fig3}d). Several striking changes as compared to the standard Hanle effect without DNP emerge for the dots investigated: First, the Hanle curve is broadened. An increase of $\Delta B$ is expected if $\tau_s^*$ decreases. This is plausible as the higher laser excitation power makes trapping of additional charges more likely, so the effective time the X$^+$ complex exists might be shortened as biexcitons are formed. Second, around zero-field we observe a pronounced W-shape as can be verified in Fig.\ref{fig:fig3}c. 

\textbf{Tilting the nuclear magnetization.} These substantial changes of the electron polarization in applied transverse magnetic fields of only a few mT are surprising. We have developed a model (see Methods for a brief and the Supplementary Information for a detailed description) based on the intricate interplay between the applied transverse field $B_Y$, the Knight field $B_e$ and the Overhauser field $B_n$ that qualitatively reproduces the main features of the measured electron depolarization curve, compare Figs.\ref{fig:fig3}c and e. \\
\indent In the simulations, we use (i) the Knight field amplitude determined in the longitudinal magnetic field measurements in Fig.\ref{fig:fig2} and (ii) $\Delta B$ that we determine in the high transverse field part of the Hanle curve were nuclear effects are negligible. We are able to reproduce qualitatively the measured nuclear spin polarization (Fig.\ref{fig:fig3}f) and electron polarization (Fig.\ref{fig:fig3}e), namely the characteristic W-shape for $|B_Y|\leq 20$~mT. At low $B_Y$ the total field $B_T$ acting on the nuclei is determined by $B_Y$ and $B_e$. This introduces a tilt in the strong nuclear field $B_n$ and hence a strong transverse field component $B_{n,Y}$ that points in the opposite direction of $B_Y$, as indicated in Fig.\ref{fig:fig3}h. The opposite sign of the transverse nuclear field $B_{n,Y}$ with respect to the applied field $B_Y$ can be directly verified in Fig.\ref{fig:fig3}f. This configuration is responsible for the sudden drop in the electron polarization. As $B_Y$ is increased and eventually fully compensates $B_{n,Y}$, the electron experiences a total field that is zero. In the experiments, we observe at this field a local maximum, as indicated in Fig.\ref{fig:fig3}c. The applied field value $B_Y$, for which this maximum occurs, serves as an experimental determination of the transverse Overhauser field in the range of 50~mT as $B_{n,Y}=-B_Y$, as reproduced by our model which allows to determine the compensation points in Fig.\ref{fig:fig3}f. The good overall agreement between the model and the experiments indicates that we included the key interactions (Knight field $B_e$, Overhauser field $B_n$, applied field $B_Y$) in our model. We can clearly see that the total nuclear field is collinear with the field experienced by the nuclei, which allows us to finely tune the orientation (tilt) of the Overhauser field in the Z-Y plane by varying the laser polarisation and $B_Y$, in stark contrast to the anomalous Hanle effect in strained InGaAs dots \cite{Krebs:2010a,Nilsson:2013a}. \\
\indent At the compensation point $B_{n,Y}=-B_Y$, the experimentally observed polarization is lower than the calculated one as we have neglected the nuclear field fluctuations $\delta B_n$ in our theoretical analysis. When the electron is not protected by a strong external and/or internal magnetic field, it will precess around $\delta B_n$, resulting in a decrease of the measured electron spin polarization \cite{Merkulov:2002a,Khaetskii:2002a}. As $B_Y$ is further increased, we enter the regime of the standard Hanle effect as the electron precesses around a strong applied magnetic field in the absence of any substantial DNP. Comparison between data and theory allows us to extract the sign of the transverse electron g-factor, which is positive. In the opposite case ($g_{e,\bot}<0$) compensation between $B_{n,Y}$ and $B_Y$ is impossible as they would have the same sign, so that the striking W-shape would not appear.\\ 
\textbf{DISCUSSION.} The combination of zero field DNP and the application of small transverse magnetic fields allows control of the longitudinal and transverse component of the magnetization of the mesoscopic nuclear spin ensemble in isolated (no wetting layer), strain free GaAs dots. All deviations from the standard Hanle curve reported here occur for applied transverse fields in the tens of mT range. On the one hand this indicates that both the Knight and the Overhauser fields are 1 to 2 orders of magnitude stronger than in the case of electrons localised at donors in bulk GaAs \cite{Paget:1977a}. On the other hand, the effects reported here occur for external magnetic fields one order of magnitude smaller compared to the very anomalous Hanle effect reported for strained InGaAs dots \cite{Krebs:2010a,Nilsson:2013a}. Our measurements confirm that strain and the associated nuclear quadrupole effects are at the origin of the intriguing observations in InGaAs dots.\\
\indent The coupled electron to nuclear spin system shows highly non-linear dynamics \cite{Urbaszek:2013a} and should be investigated in the future under (quasi-)resonant pumping conditions in order to achieve an electron spin polarization $> 50\%$. This would allow to verify the exciting predictions of phase transitions of the mesoscopic nuclear spin ensemble \cite{Kessler:2012a}, which might be enhanced due to the absence of nuclear quadrupole effects in the strain free GaAs droplet dots investigated here. Another extension of this work is the investigation of spin diffusion between two GaAs droplet dots through the AlGaAs barrier \cite{Malinowski:2001a}, in the presence \cite{Belhadj2008} or absence of a wetting layer \cite{Mano:2010a}, a comparison that cannot be made in the SK dot systems. Zero field DNP reported here is also expected to occur during all-electrical manipulations of nuclear and electron spin polarization \cite{Shiogai:2012a,Salis:2009a}, although it has not been reported yet.  \\

\textbf{METHODS: Samples and Experiments.}  The sample was grown by droplet epitaxy using a molecular
beam epitaxy system \cite{Mano:2010a,Sallen:2011a} on a GaAs(111)A substrate. 
The dots are grown on 100nm thick Al$_{0.3}$Ga$_{0.7}$As  barriers and are covered by 50nm of the same material.
In this model system dots (typical height $\simeq$2-3nm, radius $\simeq$15nm) are truly isolated as they are not connected by a two-dimensional quantum well (wetting layer) \cite{Mano:2010a}, contrary to SK dots and dots formed at quantum well interface fluctuations \cite{Gammon:1997a,Bracker:2005a}.
Single dot photoluminescence (PL) at 4K is recorded with
a home build confocal microscope with a detection spot diameter of
$\simeq 1\mu m$. The detected PL signal is dispersed by a spectrometer and detected by a Si-CCD
camera. Optical excitation is achieved by pumping the AlGaAs barrier
with a HeNe laser at 1.96~eV. Laser polarization control and PL polarization analysis is performed with Glan-Taylor polarisers and liquid crystal waveplates. The PL circular polarization degree is defined as $\rho=\frac{I_{\sigma+}-I_{\sigma-}}{I_{\sigma+}+I_{\sigma-}}$, where $I_{\sigma+}(I_{\sigma-})$ is the $\sigma^+(\sigma^-)$ polarized PL intensity.
As we excite heavy and light hole transitions in the AlGaAs barrier, 100\% laser polarization results in a maximum PL polarization $\rho_{max}=50\%$ for the X$^+$, corresponding to a maximum initial electron spin polarization of 50\%. In our experiment we find typically $\rho\simeq40\%$ for most dots at zero applied magnetic field.
The PL emission energy is determined with a precision of 1~$\mu$eV from Gaussian fits to the spectra (signal to noise ratio 10$^4$, FWHM $\simeq 40~\mu$eV limited by spectrometer resolution). The spectral precision is calibrated in experiments with a Fabry-Perot interferometer with the resolution of 0.2~$\mu$eV \cite{Belhadj:2009a}. Magnetic fields up to 9~T can be applied both parallel (Faraday geometry) and perpendicular (Voigt geometry) to the growth axis [111] that is also the angular momentum quantization axis and the light propagation axis.\\
\textbf{Model for electron depolarization in transverse magnetic field in the presence of DNP.} See Supplementary Information for a more detailed description. We assume for simplicity harmonic carrier confinement in the three spatial directions. We include in our model the possibility of a small tilt angle $\theta\leq 3^{\circ}$ in the sample holder used for transverse magnetic fields (Voigt geometry). We define the following bases $\mathcal{B}=\lbrace\bm{x},\bm{y},\bm{z}\rbrace$ where $\bm{z}$ is normal to the sample surface and $\bm{y}$ is in the plane $(O,\bm{B},\bm{z})$; $\mathcal{B'}=\lbrace\bm{x'},\bm{y'},\bm{z'}\rbrace$ where $\bm{x'}=\bm{x}$ and $\bm{y'}||\bm{B}$ obtained by rotation of $\mathcal{B}$ around axis $Ox$ by an angle $\theta=(\bm{y},\bm{B})$; $\mathbb{B}=\lbrace\bm{X},\bm{Y},\bm{Z}\rbrace$ where $\bm{Y}=\bm{y'}$ and $\bm{Z}||\bm{S}$ obtained by rotation of $\mathbb{B'}$ around axis $Oy'$ by an angle $\phi=(\bm{z'},\bm{B})$. Here $\bm{B}$ is the total field experienced by the electron of average spin $\langle S \rangle$. We neglect throughout our analysis the hyperfine coupling between nuclei and valence holes, which is about one order of magnitude weaker than for electrons \cite{Urbaszek:2013a}. The initial, photogenerated electron spin can be written as:
\begin{equation}
[S_0]_{\mathcal{B}}=S_0 \left( \begin{array}{c} 0 \\ 0 \\ 1 \end{array} \right) 
\label{eq:S0}
\end{equation}
The rotation angle $\phi$ given by the Bloch equations quantifies the precession of the electron in the total field $B=B_{Y}+B_{n,Y}$ (see scheme in Fig.\ref{fig:fig3}h) can be expressed as \cite{Abragam:1961a}:
\begin{equation}
\cos ^2\phi =\frac{\Delta B^2}{\Delta B^2+B^2_{y'}}=\frac{\Delta B^2}{\Delta B^2+(B_{Y}+B_{n,y'})^2}
\label{eq:cos2phi}
\end{equation}
where $\Delta B=\hbar/(g_{e,\bot}|\mu_B|\tau^*_s)$ and $1/\tau^*_s=1/\tau_s+1/\tau$, where $\tau$ is the radiative lifetime of the trion X$^+$. The final results of the calculation are expressed in reduced magnetic field units $\beta_{n,Y}\equiv\frac{B_{n,Y}}{\Gamma_t b_e(0)S_0}$ and $\beta_{n,Z}\equiv\frac{B_{n,Z}}{\Gamma_t b_e(0)S_0}$ and the dimensionless Hanle width (without DNP) which is $\Delta\beta\equiv\frac{B_{n,Y}}{\Gamma_t b_e(0)S_0}$. We can calculate the nuclear field components as:

\begin{eqnarray}
\beta_{n,Y}=K\lbrace \sin\theta+\cos\theta\cos\phi Im[G(\Xi)]\rbrace \label{eq:betanY}\\
\beta_{n,Z}=K \cos\theta\cos\phi \lbrace 1+Re[G(\Xi)]\rbrace \label{eq:betanZ}
\end{eqnarray}

where

\begin{eqnarray*}
\Xi=\frac{\sin\theta+i \cos\theta \cos \phi}{\beta_Y} \label{eq:xi}\\
\cos \phi \frac{\Delta \beta}{\sqrt{\Delta\beta^2+(\beta_Y+\beta_{n,Y})^2}} \label{eq:cosphi}\\
G(\Xi)=\frac{1}{\Xi}Li_{\frac{3}{2}}(-\Xi) \label{eq:gxi}\\
K=-2^{-\frac{5}{2}}\frac{\tilde{f}}{\Gamma_t}\frac{\tilde{g}_N \mu_N}{g_{e,\bot} |\mu_B|} Q N_L \label{eq:K}\\
\end{eqnarray*}

Here $Li_{\frac{3}{2}}$ is the poly-logarithmic function, $\tilde{g}_N$ is the average nuclear g-factor, $N_L$ is the number of nuclei in the dot and $Q=5$ for a GaAs dot \cite{Urbaszek:2013a}. 
$\tilde{f}$ is an average leakage factor taking into account all relevant nuclear spin relaxation mechanisms. Eq.\ref{eq:betanY} is an implicit equation, since the complex variable $\Xi$ is a function of $\beta_{n,Y}$. The constant $K$ represents the feed-back coefficient of the nuclear magnetisation of the QD on $\beta_{n,Y}$ itself. We obtain the longitudinal Overhauser field using $\beta_{n,Y}$ and $\beta_{n,Z}$:

\begin{equation}
B_{n,z}=\Gamma_t b_e(0)S_0\beta_{n,z}=\Gamma_t b_e(0)S_0 \beta_{n,Y}\sin\theta+\beta_{n,Z}\cos \theta \cos \phi 
\label{eq:Bnz}
\end{equation}

which can be compared with the measurements, see Fig.\ref{fig:fig3}(d) and (f).  
Finally, we also obtain an expression for the circular polarization degree of the $X^+$ emission:

\begin{equation}
\rho(B_Y)=-2S_z=P_0(\cos^2\theta \cos^2 \phi+\sin^2\theta)
\label{eq:Ptheo}
\end{equation}

Eq.\ref{eq:Ptheo} reproduces the measured polarization qualitatively, compare Fig.\ref{fig:fig3}c and e, using the the experimentally determined parameters $\langle B_e \rangle_{QD}^{max}=18$~mT, $g_{e,\bot}=0.78$, $\delta_n=12 \mu eV$ and $\Delta B = 80~$mT. The following parameters have been adjusted within plausible bounds to reproduce closely the experiments: $N_L=40000$, $\tilde{f}=0.22$, $\theta = 0.5^{\circ}$ and $\Gamma_t=0.3$.

\textbf{Acknowledgements} This work was partially funded by ITN SpinOptronics, ANR QUAMOS and ERC StG OptoDNPcontrol.\\


\end{document}